%% file: eptcs10.tex
\nonstopmode
\documentclass[a4paper,copyright,creativecommons]{eptcs} % ,submitted
\providecommand{\event}{PAR 2010}
\usepackage{breakurl} % recommended by eptcs style
\usepackage{ifthen}
\usepackage{amsmath}
\usepackage{amssymb}
\usepackage{stmaryrd}

\ifdefined\LONGVERSION
  \relax
\else
\newcommand{\LONGVERSION}[1]{}

\fi

\input{macros}

\def\squareforqed{\ensuremath{\Box}}
\def\qed{\ifmmode\squareforqed\else{\unskip\nobreak\hfil
\penalty50\hskip1em\null\nobreak\hfil\squareforqed
\parfillskip=0pt\finalhyphendemerits=0\endgraf}\fi}

\newenvironment{proof*}[1][]{\noindent\ifthenelse{\equal{#1}{}}{{\it
      Proof.}}{{\it Proof #1.}}\hspace{2ex}}{\bigskip}

\newcommand{\Size}{\mathsf{Size}}

\newcommand{\Nat}{\mathsf{Nat}}
\newcommand{\tzero}{\mathsf{zero}}
\newcommand{\tsucc}{\mathsf{succ}}

\renewenvironment{samepage}{}{}
\newenvironment{code}{\begin{samepage}\begin{quote}}{\end{quote}\end{samepage}}

\title{MiniAgda: Integrating Sized and Dependent Types}
\def\titlerunning{MiniAgda: Integrating Sized and Dependent Types}
\author{Andreas Abel
  \institute{
    Department of Computer Science \\
    Ludwig-Maximilians-University Munich, Germany
    \thanks{Part of this research was carried out
      as invited researcher (Oct 2009 -- Mar 2010) in the PI.R2 project
      team of INRIA Rocquencourt at the PPS lab, Universit\'e Paris VII.
    }%
  }%
  \email{andreas.abel@ifi.lmu.de}%
}
\def\authorrunning{Andreas Abel}
\begin{document}
\maketitle

% \begin{abstract}
%   We present MiniAgda, a core dependently typed functional language.
%   MiniAgda is total, \ie, each recursive function is terminating and
%   each corecursive function is productive.  Termination and
%   productivity is ensured by the type system using sized types.  Sizes
%   are explicit in MiniAgda, they must be supplied in the definition of
%   inductive and coinductive data types and in the types of recursive
%   and corecursive functions.  
%   % Sizes can partially be reconstructed by a constraint solver.  
%   MiniAgda is intended as a test bed for
%   termination checking using sized types.  This paper only provides an
%   informal introduction to MiniAgda, without any meta-theoretical results.
% \end{abstract}

\begin{abstract}\noindent
Sized types are a modular and theoretically well-understood tool for
checking termination of recursive and productivity of corecursive
definitions.  The essential idea is to track structural descent and
guardedness in the type system to make termination checking robust and
suitable for strong abstractions like higher-order functions and
polymorphism.
To study the application of sized types to proof assistants and
programming languages based on dependent type theory, we have
implemented a core language, MiniAgda, with explicit handling of
sizes.  New considerations were necessary to soundly integrate sized
types with dependencies and pattern matching, which was made possible
by concepts such as inaccessible patterns and parametric function
spaces.  This paper provides an introduction to MiniAgda by example
and informal explanations of the underlying principles.
\end{abstract}

\section{Introduction}

In the dependent type theories underlying the programming and proof
languages of Coq \cite{inria:coq82}, Agda \cite{norell:PhD}, and
Epigram \cite{mcBrideMcKinna:view},
% \cite{chapmanAltenkirchMcBride:epigram2}, 
all programs need to be
total to maintain logical consistency.  This means that some analysis
is required that ensures that all functions defined by recursion
over inductive types terminate and all functions defined by
corecursion into a coinductive type are productive, \ie, always yield
the next piece of the output in finite time.  The currently
implemented termination analyses are based on the untyped structural
ordering:  In case of Coq, it is the guard condition
\cite{gimenez:guardeddefinitions}, and in case of Agda, the
$\mathsf{foetus}$ termination checker \cite{abelalti:predicative} with a
recent extension to size-change termination
\cite{jones:sizeChange,wahlstedt:PhD}.  The untyped approaches have some
shortcomings, including the sensitivity of the termination checker to
syntactical reformulations of the programs, and a lack of means to
propagate size information through function calls.  

As alternative to untyped termination checking, type-based methods
have been proposed \cite{gimenez:typeBased,xi:terminationHOSC,blanqui:rta04,abel:lmcs07}.
The common idea is to annotate data types with a size index which is
either the precise size of elements in this slice of the data type or
an upper bound on the size.  For recursive calls it is checked that
the sizes decrease, which by well-foundedness entails termination.
Sized types provide a very principled approach to termination, yet the
need to integrate sizes into the type system means that an
implementation of sized types touches the very core of type theories.
Most theoretical works have confined themselves to study sized types
in a simply-typed or polymorphic setting, exceptions are
Blanqui's work \cite{blanqui:rta04,blanqui:csl05} 
which adds sizes to the Calculus
of Algebraic Constructions ($\CACSA$), and $\CIChat$
\cite{bartheGregoirePastawski:lpar06}, which extends the Calculus of
Inductive Constructions in a similar fashion.   
To this day, no mature implementation of
sized types exists.

The language MiniAgda, which shall be presented in this article, is an
implementation of a dependently typed core language with sized types.
Developed by the author and Karl Mehltretter
\cite{mehltretter:diploma}, it serves to explore the interaction of
sized types with the other features of a dependently typed functional
language which include pattern matching and large eliminations.  A
fundamental design choice of MiniAgda was that sizes are explicit:
they are index arguments of data types, abstraction over sizes is
ordinary lambda-abstraction, and instantiation of a size is ordinary
application.  One of the main lessons learned during thinking about
the relationship of sized types and dependent types is that size arguments
to functions are parametric, i.e., functions can never depend on
sizes, sizes only serve to ensure termination and should be erased
during compilation.  Parametric functions are typed via a variation
of Mishra-Linger and Sheard's Erasure Pure Type Systems (EPTS)
\cite{mishraLingerSheard:fossacs08} which rest on
Miquel's \cite{miquel:tlca01} ``implicit
quantification''.\footnote{My original inspiration was 
Bernardo and Barras' Church-style version of implicit quantification
\cite{barrasBernardo:fossacs08} which is very similar to EPTS.}

In the following we shall walk through the features of MiniAgda by
example and explain the taken design choices.  Formalization and
meta-theoretical study of MiniAgda is work in progress.

\section{Sized Types for Termination}

Type-based termination
\cite{hughesParetoSabry:popl96,xi:terminationHOSC,gimenez:typeBased,blanqui:rta04,abel:PhD} 
rests on two simple principles: 

\begin{enumerate}
\item Attaching a size $i$ to each inductive type $D$.
\item Checking that sizes decrease in recursive calls.
\end{enumerate}

\subsection{Attaching Sizes to Inductive Types}

We attach a size $i$ to each inductive type $D$, yielding a sized
type $D^i$ which contains only those elements of $D$ whose height is
below $i$.  For the calculation of the height of an element $d$ of $D$
consider $d$ as a tree, where the constructors of $D$ count as nodes.
In the case of $\Nat$ we have a linear tree with inner $\tsucc$-nodes and a
$\tzero$-leaf; the size of an element $\Nat$ is the number of
constructors, which is the value of the number plus 1.  In the case of
$\List$, we also have linear trees, since in $\tcons\;a\;\vas$ only
$\vas$ of type $\List$ counts as subtree (even if $a$ happens to be a
list also).  The height of a list is its length plus one, since
$\tnil$ is also counted.  For the type of trees the size corresponds
to the usual concept of tree height.  In the case of infinitely
branching trees, the height can be transfinite; this is why sizes are
ordinals in general and not natural numbers.  However, for the
examples we present in the following, sizes will not exceed $\omega$,
so one may think about them as natural numbers for now.

Following the principle described above,
sized natural numbers are given by the constructors:
\[
\begin{array}{lll}
  \tzero & : & \forall i.\; \Nat^{i + 1} \\
  \tsucc & : & \forall i.\; \Nat^i \to \Nat^{i + 1} \\
\end{array}
\]
The first design choice is how to represent size expressions $i$ and $i+1$,
sized types $\Nat^i$ and size polymorphism $\forall i$.  In absence of
dependent types, like in the work of Pareto \etal\
\cite{hughesParetoSabry:popl96} and Barthe \etal\
\cite{gimenez:typeBased}, size
expressions and sized types need to be special constructs, or, 
in a $\Fomega$-like system, are modeled as type
constructors \cite{abel:PhD}.  In a dependently typed setting, we can
model sizes as a type $\Size$ with a successor operation $\dollar : \Size
\to \Size$, sized types just as members of $\Size \to \Set$ (where
$\Set$ is the type of small types), and size quantification just as
dependent function space $(i : \Size) \to C\;i$.  This way, we can
pass sizes as ordinary arguments to functions and avoid
special syntactic constructs.

In MiniAgda, a sized type is declared using the \verb|sized data|
syntax. 
\begin{code}
\begin{verbatim}
sized data SNat : Size -> Set
{ zero : (i : Size) -> SNat ($ i)
; succ : (i : Size) -> SNat i -> SNat ($ i)
}
\end{verbatim}
\end{code}
This declares an inductive family $\SNat$ indexed by a size with
two explicitly sized constructors.  The system checks that the type
of $\SNat$ is a function type with domain $\Size$, and the size index
needs to be the first index of the family.  The constructor
declarations must start with a size quantification $(i : \Size) \to
\dots$.  Each recursive occurrence of $\SNat$ needs to be labeled with
size $i$ and the target with size $\dollar i$, to comply with the
concept of size as tree height as described above.  The size
assignment to constructors is completely mechanical and could be done
automatically.  Indeed, in a mature system like Agda or Coq, size
assignment should be automatic, adding size information as hidden
arguments.  However, in MiniAgda, which lacks a reconstruction
facility for omitted hidden arguments, we chose to be fully explicit.
In our case, the advantage of explicit size assignment over automatic
assignment is ``what you see is what you get'', sized constructors are used
as they are declared.  

As a first exercise, we write a ``plus 2'' function for sized natural numbers.
\begin{code}
\begin{verbatim}
let inc2 : (i : Size) -> SNat i -> SNat ($$ i)
         = \ i -> \ n -> succ ($ i) (succ i n) 
\end{verbatim}
\end{code}
In this non-recursive definition $\lambda i \lambda n.\;
\tsucc\;(\dollar i)\;(\tsucc\;i\;n)$, 
we first abstract over the size $i$ of the argument $n$ using an
ordinary $\lambda$, and supply the size arguments to $\tsucc$
explicitly.
To define a predecessor function on $\SNat$, we use pattern matching.
\begin{samepage}
\begin{code}
\begin{verbatim}
fun pred : (i : Size) -> SNat ($$ i) -> SNat ($ i)
{ pred i (succ .($ i) n) = n
; pred i (zero .($ i))   = zero i
}
\end{verbatim}
\end{code}  
\end{samepage}
To avoid non-linear left-hand sides of pattern matching equations, 
MiniAgda has 
\emph{inaccessible patterns} \cite{bradyMcBrideMcKinna:types03}, 
also called \emph{dot patterns}.
A dot pattern can be an arbitrary term, in our case it is the size
expression $\$ i$.  A dot pattern does not bind any variables, and during
evaluation dot patterns are not matched against.  The dot pattern
basically says ``at this point, the only possible term is $\$ i$''.  
% To define a predecessor function on $\SNat$, we use pattern matching.
% \begin{code}
% \begin{verbatim}
% data MaybeNat (i : Size) : Set
% { nothing : MaybeNat i
% ; just    : SNat i -> MaybeNat i
% }

% fun pred : (i : Size) -> SNat ($ i) -> MaybeNat i
% { pred i (succ .i n) = just i n
% ; pred i (zero .i)   = nothing i
% }
% \end{verbatim}
% \end{code}
% To avoid non-linear left-hand sides of pattern matching equations, 
% MiniAgda has \emph{inaccessible patterns}
% \cite{bradyMcBrideMcKinna:types03}, also called \emph{dot patterns}.
% A dot pattern can be an arbitrary term, in our case it is just
% variable $i$.  A dot pattern does not bind any variables, and during
% evaluation dot patterns are not matched against.  The dot pattern
% basically says ``at this point, the only possible term is $i$''.  In our
% example, it does not matter where we place the dot, we could also have
% written:
% \begin{code}
% \begin{verbatim}
% { pred .i (succ i n) = just i n
% ; pred .i (zero i)   = nothing i
% }
% \end{verbatim}
% \end{code}
 
To ensure totality and consistency, pattern matching must be complete.
Currently, MiniAgda does not check completeness; the purpose of
MiniAgda is to research sized types for integration into mature
systems such as Agda and Coq, which already have completeness checks.

\subsection{Functions are Parametric in Size Arguments}

Sizes are a means to ensure termination, so they are static
information for the type checker, and they should not be present at
run-time.  In particular, the result of a function should never be
influenced by its size arguments, and we should not be allowed to
pattern match on sizes.  Only then it is safe to erase them during
compilation.   However, the \emph{type} of a function does very well
depend on size; while the first argument to $\tpred$ is irrelevant, so
we have $\tpred~i~n = \tpred~j~n$ for all sizes $i,j$, the types
$\SNat\;(\dollar i)$ and $\SNat\;i$ are certainly different.  The
concept of ``irrelevant in the term but not in the type'' is nicely
captured by Miquel's intersection types \cite{miquel:tlca01} which
have been implemented by Barras \cite{barrasBernardo:fossacs08} as a
variant $ICC^*$ of Coq and studied in Mishra-Linger's thesis
\cite{mishraLinger:PhD}.   
Intersection types, which we write as $[x :
A] \to B$ and would like to call \emph{parametric function types}, 
are a generalization of polymorphism $\forall X.B$ in System~F.
Polymorphic functions also do not depend on their type arguments, but
their types do.

Using parametric function types, we can refine our definition of
$\SNat$ as follows.
\begin{code}
\begin{verbatim}
sized data SNat : Size -> Set
{ zero : [i : Size] -> SNat ($ i)
; succ : [i : Size] -> SNat i -> SNat ($ i)
}
\end{verbatim}
\end{code}
We have now expressed that the size argument to the constructors is
irrelevant, so they can be safely erased during compilation.  Size
arguments can \emph{always} be irrelevant in programs.  The type of
predecessor is consequently refined as:
\begin{samepage}
\begin{quote}
\begin{code}
\begin{verbatim}
fun pred : [i : Size] -> SNat ($$ i) -> SNat ($ i)
{ pred i (succ .($ i) n) = n
; pred i (zero .($ i))   = zero i
}
\end{verbatim}
\end{code}
\end{quote}
\end{samepage}
Note that we do bind variable $i$ to the irrelevant size argument, but
the type system of MiniAgda ensures that it only appears on the right
hand side in irrelevant positions, i.e., in arguments to parametric
functions.  
\begin{code}
\begin{verbatim}
let inc2 : [i : Size] -> SNat i -> SNat ($$ i)
         = \ i -> \ n -> succ ($ i) (succ i n) 
\end{verbatim}
\end{code}
Here, $i$ is used as an argument to the parametric $\tsucc$, which is
fine.  We would get a type error with the non-parametric definition of
$\tsucc$:
\begin{code}
\begin{verbatim}
succ : (i : Size) -> SNat i -> SNat ($ i)
\end{verbatim}
\end{code}
The precise typing rules for parametric functions have been formulated
by Mishra-Linger and Sheard~\cite{mishraLingerSheard:fossacs08}.  For our
purposes, the informal explanation we have just given is sufficient.

\subsection{Tracking Termination and Size Preservation}

Consider the following implementation of subtraction.  Here, we use
the infinity size $\infty$, in concrete syntax $\#$.  An element of
sized type $D\;\infty$ is one without known height bound.  This means
that for sized inductive types $D$ we have the subtyping relationships
$D\;i \leq D\;(\$ i) \leq \dots \leq D\;\infty$.

Also, we introduce \emph{size patterns} $i > j$ to match a constructor 
$c : [j :  \Size] \to \dots \to D\;(\$ j)$ against an argument of type
$D\;i$.  The size pattern $i > j$ binds size variable $j$ and
remembers that $j$ is smaller than size variable $i$.
The full pattern we write is $c\;(i > j)\;\vec p$ and the match
introduces a size constraint $i > j$ into the type checking
environment which is available to check the remaining patterns and the
right hand side of the clause.
\begin{code}
\begin{verbatim}
fun minus : [i : Size] -> SNat i -> SNat # -> SNat i
{ minus i (zero (i > j))    y          = zero j
; minus i  x               (zero .#)   = x
; minus i (succ (i > j) x) (succ .# y) = minus j x y   
}
\end{verbatim}
\end{code}
The last line contains one recursive call $\tminus~j~x~y$ with a
descent in all three arguments:  the size decreases ($j < i$)
and both $\SNat$-arguments are structurally smaller ($x < \tsucc~j~x$
and $y < \tsucc~\infty~y$).  Thus, termination can be certified by descent
on any of the three arguments. 

The type of $\tminus$ expresses that $\tminus$ takes an argument $x$
of size at most $i$ and an argument $y$ of arbitrary size
and returns a result of size at most $i$.  Thus, sizes
can be used to certify that the output is bound in size by one of the
inputs.  In our case, the sized type expresses that ``subtracting does
not increase a number''.  In the last line, we return $\tminus~j~x~y$
which is of size $j$, but the type requires us to return something of
size $i$.  Since sizes are upper bounds, by the subtyping of
MiniAgda we have $\SNat\;j \leq \SNat\;i$, so this example passes
the type checker.  To see that the first clause is well-typed, observe
that $\tzero\;j : \SNat\;(\$ j)$, and since $j < i$ entails $\$ j \leq
i$, we have $\tzero\;j : \SNat\;i$ by subtyping.
Note that in MiniAgda we cannot express that the result of 
$\tminus\;i\;x\;(\tsucc\;\infty\;y)$ is strictly smaller in size than~$x$.  
% However, $\tminus\;(\$ i)\;(\tpred\;i\;x)\;y : \SNat\;(\$ i)$ is strictly
% smaller than $x : \SNat\;(\$\$ i)$.

Using the bound on the output size of $\tminus$ we can certify
termination of Euclidean division.  A call to $\tdiv\;i\;x\;y$
computes $\lceil x/(y+1) \rceil$.
\begin{code}
\begin{verbatim}
fun div : [i : Size] -> SNat i -> SNat # -> SNat i
{ div i (zero (i > j))   y = zero j 
; div i (succ (i > j) x) y = succ j (div j (minus j x y) y)
}
\end{verbatim}
\end{code}
In the last line, since $x$ is bounded by size $j$, so is
$\tminus\;j\;x\;y$, hence, the recursive call to $\tdiv$ can happen at
size $j$ which is smaller than $i$.
That tracking a size bound comes for free with the type
system is a major advantage of sized types over untyped termination
approaches.

\subsection{Interleaving Inductive Types}

Another advantage of sized types is that they scale very well to
higher-order constructions.  In the following, we present an
implementation of $\tmap$ for rose trees to demonstrate this feature.

If we do not require sizes, we can also define ordinary data types in
MiniAgda.  In a mature system, all data definitions would look
``ordinary'' but be sized behind the veil, so there would not be a need
for separate syntactic constructs.
\begin{code}
\begin{verbatim}
data List ++(A : Set) : Set
{ nil  : List A
; cons : A -> List A -> List A
}
\end{verbatim}
\end{code}
$\List$ is a parametric data type, and we declare that it is
strictly positive in its parameter $A$ by prefixing that parameter by
two $+$ signs.  This means in particular that $\List$ is a
monotone type valued function, i.e., for $A \leq B$ we have $\List\;A
\leq \List\;B$; this property is beneficial for subtype checking.  
Internally, the constructors of $\List$ are stored as follows:
\[
\begin{array}{lll}
  \tnil  & : & [A : \Set] \to \List\;A \\
  \tcons & : & [A : \Set] \to A \to \List\;A \to \List\;A \\ 
\end{array}
\]
Each parameter of the data type becomes a parametric argument of the
data constructors.  \footnote{This is a consequent analysis of data type
parameters, compare this to Coq's handling of parameters which
requires parameters in constructors to be absent in patterns but
present in expressions.}

The $\tmap$ function for lists is parametric in types $A$ and $B$.
Observe the dot patterns in the matching against $\tnil$ and $\tcons$!
\begin{code}
\begin{verbatim}
fun mapList : [A : Set] -> [B : Set] -> (A -> B) -> List A -> List B
{ mapList A B f (nil .A) = nil B
; mapList A B f (cons .A a as) = cons B (f a) (mapList A B f as)
}
\end{verbatim}
\end{code}

Rose trees are finitely branching node-labeled trees, where the
collection of subtrees is organized as a list.
\begin{code}
\begin{verbatim}
sized data Rose ++(A : Set) : Size -> Set
{ rose : [i : Size] -> A -> List (Rose A i) -> Rose A ($ i) 
}
\end{verbatim}
\end{code}
MiniAgda checks data type definitions for strict positivity to avoid
well-known paradoxes.  In this case, it needs to ensure that $\Rose$
appears strictly positively in the type $\List\;(\Rose\;A\;i)$.  The
test is trivially successful since we have declared $A$ to appear only
strictly positively in $\List\;A$ above.  
% Positivity annotations allow
% for a trivial implementation of the positivity test, unlike the
% current implementations in Agda and Coq which require unfolding of
% definitions during the positivity check.
Also, the type argument $A$ appears strictly positively in $\Rose$ itself
since strictly positive type functions compose.  So the positivity
annotation of $\Rose$ is sound, and we could continue and define trees
branching over roses etc.

A feature of sized types is that the map function for roses can be
defined naturally using the map function for lists:
\begin{code}
\begin{verbatim}
fun mapRose : [A : Set] -> [B : Set] -> (A -> B) -> 
              [i : Size] -> Rose A i -> Rose B i
{ mapRose A B f i (rose .A (i > j) a rs) = 
  rose B j (f a) (mapList (Rose A j) (Rose B j) (mapRose A B f j) rs)
}
\end{verbatim}
\end{code}
Note that the recursive call $\tmapRose~A~B~f~j$ is underapplied, the
actual rose argument is missing.  But since the size $j$ of the rose
argument is present, we can verify termination.\footnote{Sized types even allow
nested recursive calls \cite{abel:scp09}.}  This is usually a severe
problem for untyped termination checkers.  The current solution in Coq
requires $\Rose$ and $\List$ to be defined mutually, destroying
important modularity in library development.

An alternative to sized types is the manual indexing of roses by their
height as \emph{natural number}.  This can be done within the scope of
ordinary dependent types.  Yet we miss the comfortable subtyping for sized
types.  The height of a rose has to be
computed manually using a maximum over a list, and the resulting
programs will involve many casts to type-check.

\section{Coinductive Types and Corecursion}

Coinductive types admit potentially infinite inhabitants, the most
basic and most prominent example being streams.
\begin{code}
\begin{verbatim}
codata Stream ++(A : Set) : Set 
{ cons : A -> Stream A -> Stream A
}
\end{verbatim}
\end{code}
An element of $\Stream\;A$ is an infinite succession of elements of
$A$ knitted together by the $\tcons$ coconstructor.  Clearly, a
function producing a stream cannot terminate, since it is supposed to
produce an infinite amount of data.  Instead of termination we require
\emph{productivity} \cite{coquand:infiniteobjects} which means that
always the next portion of the stream can be produced in finite time.
A simple criterion for productivity which can be checked
syntactically is guardedness \cite{gimenez:guardeddefinitions}:  
In the function definition, we require the right hand side to be a
coconstructor or a sequence of such, and
each recursive call to be directly under, \ie, \emph{guarded} by this
coconstructor.  This condition ensures that the recursive computation
only continues after some initial piece of the result has been produced.
For example, consider $\trepeat\;A\;a$ which produces an infinite
stream of $a$s. 
\begin{code}
\begin{verbatim}
cofun repeat : [A : Set] -> (a : A) -> Stream A
{ repeat A a = cons A a (repeat A a)
}
\end{verbatim}
\end{code}
The recursive call to $\trepeat$ is directly under the coconstructor
$\tcons$, so the guard condition is satisfied.  The ``directly under''
is crucial, if we write $\tcons\;A\;a\;(f\;(\trepeat\;A\;a))$ instead,
then productivity is no longer clear, but depends on $f$.  If $f$ is a
stream destructor like the $\ttail$ function, which discards the first
element of the stream and returns the rest, then we do not obtain a
productive function.  Indeed, for this definition of $\trepeat$, the
expression $\ttail\;(\trepeat\;A\;a)$ reduces to
itself after one unfolding of the recursion, so the next element of
the stream can never be computed.  On the other hand, if $f$ is a
stream-constructing function or a depth-preserving function like the
identity, then the new $\trepeat$ is productive.  Yet to be on the safe
side, the syntactic guard condition as described by Coquand
\cite{coquand:infiniteobjects} and implemented in Coq needs to
reject the definition nevertheless.  In practice this is
unsatisfactory, and some workarounds have been suggested, like
representing streams as functions over the natural numbers
\cite{bertotKomendantskaya:types08} or 
defining a mixed inductive-coinductive type of streams
with an additional constructor $c_f$
%adding a coconstructor $c_f$ to the coinductive type
which will be evaluated to $f$ in a second step
\cite{danielsson:beating}.

\subsection{Tracking Guardedness with Sized Types}

Using sized coinductive types we can keep track in the type system
whether a function is stream destructing, stream constructing or depth
preserving \cite{hughesParetoSabry:popl96}.  Thus, sized types can
offer a systematic solution to the
``guardedness-mediated-by-functions'' problem.

Sized coinductive type definitions look very similar to their inductive
counterpart: The rules to annotate the recursive occurrences of the
coinductive type in the types of the coconstructors are identical to
the rules for sized constructors.
\begin{code}
\begin{verbatim}
sized codata Stream ++(A : Set) : Size -> Set 
{ cons : [i : Size] -> A -> Stream A i -> Stream A ($ i)
}

fun head : [A : Set] -> [i : Size] -> Stream A ($ i) -> A 
{ head A i (cons .A .i a as) = a
}

fun tail : [A : Set] -> [i : Size] -> Stream A ($ i) -> Stream A i
{ tail A i (cons .A .i a as) = as
}
\end{verbatim}
\end{code}
What is counted by the size $i$ of a Stream is a lower bound on the
number of coconstructors or guards, which we call the \emph{depth} of
the stream.  A fully constructed stream will
always have size $\infty$, but during the construction of the stream
we reason with approximations, \ie, streams which have depth $i$ for
some arbitrary $i$.  This is dual to the definition of recursive
functions over inductive types.  Once they are fully defined, they
handle trees of arbitrary height, \ie, size $\infty$, but to perceive
their termination we assume during their construction that they can only
handle trees up to size $i$, and derive from this that they handle
also trees up to size $i+1$.

Using sized types, the definition of $\trepeat$ looks as follows:
\begin{code}
\begin{verbatim}
cofun repeat : [A : Set] -> (a : A) -> [i : Size] -> Stream A i
{ repeat A a ($ i) = cons A i a (repeat A a i)
}
\end{verbatim}
\end{code}
Assuming that $\trepeat\;A\;a\;i$ produces a well-defined stream of
depth $i$, we have to show that $\trepeat\;A\;a\;(\$ i)$ produces a
stream of depth $i+1$.  This is immediate since $\tcons$ increases the
depth by one.
Technically, we have used a \emph{successor pattern} $(\$ i)$ on the
left hand side.  Matching on a size argument seems to violate the
principle that sizes are irrelevant for computation.  Also, not every
size is a successor, so the matching seems incomplete.
% However, since sizes are irrelevant we should not be able to
% match on them, or?  And just a successor pattern should not cover all
% possible cases, or?  
However, the match is justified by the greatest fixed-point semantics
of coinductive types, as we will explain in the following.  

In the semantics, we construct the
approximation $\Stream\;A\;i$ of the full type of streams over $A$
by an induction on $i \in \{0,1,\dots,\omega\}$.
\[
\begin{array}{lll}
  \Stream\;A\;0 & = & \top \\
  \Stream\;A\;(i+1) & = & \{ \tcons\;A\;i\;a\;s \mid a \in A \mand 
     s \in \Stream\;A\;i \} \\
  \Stream\;A\;\omega & = & \bigcap_{i < \omega} \Stream\;A\;i \\
\end{array}
\]
This is an approximation from above: we start at $\top$, the biggest
set of our semantics, \eg, the set of all terms.  A
$\Stream\;A\;0$ can be arbitrary, there is no guarantee of what will
happen if we try to look at its first element.  To be on the safe
side, we have to assume that
taking the head or tail of such a ``stream'' will diverge.  Now each
constructor increases the depth of the stream by one, we obtain
$\Stream\;A\;1$, $\Stream\;A\;2$ etc.  An element of $\Stream\;A\;i$
can be unrolled (at least) $i$ times, \ie, we can take the tail $i$
times without risking divergence.  Finally, the limit
$\Stream\;A\;\omega$ is
defined as the intersection of all $\Stream\;A\;i$.  Such a stream is
completely defined, and we unroll it arbitrarily often.

Whenever we have to construct a $\Stream\;A\;i$ for $i$ a size
variable, we can match size $i$ against a successor pattern $(\$ j)$.
Clearly, if $i$ stands for a successor size $n + 1$, then the match
succeeds with $j = n$.  
But also, if $i$ stands for size $\omega = \infty$ which is the
closure ordinal for all coinductive types then the match
succeeds with $j = \infty$ (there is a catch, see Section~\ref{sec:adm}).  
Now if $i$ stands for
$0$, we have to produce something in $\Stream\;A\;0$, meaning that we
can return anything; we have no specific obligation.  In summary, if
we produce something in $\Stream\;A\;i$, matching $i$ against a
successor pattern is a complete match.  Further, since there is only
one case (successor), the match is irrelevant and we maintain the
property that sizes do not matter computationally.  Size matching on
$i$ can be generalized to the case that we have to produce something
in $(\vec x : \vec B) \to \Stream\;A\;i$.

\subsection{Depth-preserving Functions and Mediated Guardedness}

Using the successor pattern, we can define $\tmap$ as a
depth-preserving corecursive function on streams:
\begin{code}
\begin{verbatim}
cofun map : [A : Set] -> [B : Set] -> [i : Size] -> 
            (A -> B) -> Stream A i -> Stream B i 
{ map A B ($ i) f (cons .A .i x xs) = cons B i (f x) (map A B i f xs)
}
\end{verbatim}
\end{code}
By matching first on the size $i$ (which is legal since 
we construct a $(A \to B) \to \Stream\;A\;i \to \Stream\;B\;i$
at this point), the
last argument gets type $\Stream\;A\;(\$ i)$, so we can match on it,
the only case being $\tcons\;.A\;.i\;x\;\vxs$.  Note that unlike for
inductive types we cannot
match on $\Stream\;A\;i$ since it might be a totally undefined
stream.  Technically, the matching on $\Stream\;A\;i$ is prevented in
MiniAgda by requiring that the size argument of a coconstructor
pattern must be a dot pattern. 

Similarly to $\tmap$ we define $\tmerge$ for streams.  Its type
expresses that the output stream has at least the same depth as both input
streams have.  The more precise information that its depth is the sum
of the depths of the input streams is not expressible in the size
language of MiniAgda.  The information loss is substantiated by the
use of subtyping $\Stream\;A\;(\$ i) \leq \Stream\;A\;i$ in the
recursive calls to $\tmerge$.
\begin{code}
\begin{verbatim}
cofun merge : [i : Size] -> Stream Nat i -> Stream Nat i -> Stream Nat i
{ merge ($ i) (cons .Nat .i x xs) (cons .Nat .i y ys) = 
      leq x y (Stream Nat ($ i))
         (cons Nat i x (merge i xs (cons Nat i y ys)))
         (cons Nat i y (merge i (cons Nat i x xs) ys))     
}
\end{verbatim}
\end{code}
The code of $\tmerge$ uses an auxiliary function $\tleq$ that compares
to natural numbers and returns a Church-encoded Boolean
(continuation-passing style).  In other
words, it is a fusion of a comparison and an if-then-else.
\begin{code}
\begin{verbatim}
fun leq : Nat -> Nat -> [C : Set] -> C -> C -> C
{ leq  zero     y       C t f = t
; leq (succ x)  zero    C t f = f
; leq (succ x) (succ y) C t f = leq x y C t f 
}
\end{verbatim}
\end{code}
An example where corecursion goes through another function is the
Hamming function which produces a stream of all the composites of two
and three in order.  In this case, the guarding coconstructor and the
recursive calls are separated by applications of the depth-preserving 
$\tmap$ and $\tmerge$.
\begin{code}
\begin{verbatim}
let double : Nat -> Nat = ...
let triple : Nat -> Nat = ...

cofun ham : [i : Size] -> Stream Nat i
{ ham ($ i) = cons Nat i (succ zero) 
                (merge i (map Nat Nat i double (ham i)) 
                         (map Nat Nat i triple (ham i)))
}
\end{verbatim}
\end{code}
Summing up, sized types offer an intuitive and comfortable way to
track guardedness through function calls and overcome the limitations
of syntactical guardedness checks.  The technical solutions to
integrate pattern matching with explicitly sized coconstructors are
the successor pattern for sizes and the restriction of size matching
to dot patterns in coconstructors.

% BEGIN LONG %%%%%%%%%%%%%%%%%%%%%%%%%%%%%%%%%%%%%%%%%%%%%%%%%%%%%%%%%
\LONGVERSION{

\subsection{Deep Guardedness}

The stream of Fibonacci numbers is given by the recursive equation
$\tfib = (0, (1, \tfib) + \tfib)$ with $+$ being point-wise addition of
infinite tuples.  This can be implemented directly in MiniAgda.
\begin{code}
\begin{verbatim}
cofun adds : [i : Size] -> Stream Nat i -> Stream Nat i -> Stream Nat i 
{ adds ($ i) (cons .Nat .i a as) (cons .Nat .i b bs) = 
    cons Nat i (add a b) (adds i as bs)
}

cofun fib : [i : Size] -> Stream Nat i
{  fib ($ i) = cons Nat i 0 (adds (cons Nat i 1 (fib i)) (fib i))
}
\end{verbatim}
\end{code}
The alternative equation $\tfib = (0, 1, \tfib + \ttail\;\tfib)$ is
equivalent, but harder to check for guardedness.  Intuitively, the
recursive calls to $\tfib$ are under two guards, so taking away one
guard via the $\ttail$ function still leaves one guard, and
productivity ensues.  We would like to be able to write this
equation in MiniAgda.
\begin{code}
\begin{verbatim}
cofun fib : [i : Size] -> Stream Nat i
{ fib ? = cons Nat ? 0 
    (cons Nat ? 1 (adds ? (fib ?) (tail Nat ? (fib ?))))
}
\end{verbatim}
\end{code}
The question is how to fill in the missing sizes.  A first idea is to
extend successor patterns to multi-successor patterns:
\begin{code}
\begin{verbatim}
cofun fib : [i : Size] -> Stream Nat i
{ fib ($$ i) = cons Nat ($ i) 0 
    (cons Nat i 1 (adds i (fib i) (tail Nat i (fib ($ i)))))
} 
\end{verbatim}
\end{code}
While fine in this case, multi-successor patterns are unsound in the
general case.  There is no semantic justification for such a pattern,
and it is easy to write non-productive functions:
\begin{code}
\begin{verbatim}
cofun bad : [i : Size] -> Stream Nat i
{ bad ($$ i) = cons Nat ($ i) (head Nat i (bad ($ i))) 
   (cons Nat i 1 (tail Nat i (bad ($ i))))
}
\end{verbatim}
\end{code}
While the second recursive call to $\tbad$ is still sufficiently
guarded, the first recursive call is not, but it goes unnoticed by the
type system.  Taking the head of $\tbad$ loops.

Thus, we need to stick to the simple successor pattern.  Starting to
write the definition,
\begin{code}
\begin{verbatim}
cofun fib : [i : Size] -> Stream Nat i
{ fib ($ i) = cons Nat i 0 ? 
}
\end{verbatim} %
\end{code}
we realize that the expression to replace the question mark has type
$\Stream\;\Nat\;i$ which is a coinductive type of size $i$, a
variable.  At this point, we could match $i$ against successor pattern
$\$ j$, yet we are not on the left hand side of the equation but 
within a term.  To enable size matches also on the right hand side, a
special case construct is available in MiniAgda:
\[
\rux{\Gamma, j < i \der \subst {\$ j} i t : \subst{\$ j} i C \qquad
     \Gamma \der C = (\vec x : \vec B) \to \Stream\;A\;i 
   }{\Gamma \der \tcase\;i\;\tof\;(\$ j) \to t : C
   }
\]
TODO: NEED A SIDECONDITION ABOUT CONTINUITY!

The soundness of this typing rule is analogous to the soundness of
successor patterns we argued above.  It allows us to complete the
definition of $\tfib$.
\begin{code}
\begin{verbatim}
cofun fib : [i : Size] -> Stream Nat i
{ fib ($ i) = cons Nat i 0 (case i 
    { ($ j) -> cons Nat j 1 (adds j (fib j) (tail Nat j (fib i))) })
} 
\end{verbatim}
\end{code}
Checking the case branch against type $\Stream\;\Nat\;(\$ j)$
succeeds with $\$ j \leq i$ which entails
$\Stream\;\Nat\;i \leq \Stream\;\Nat\;(\$\;j)$.

}
% END LONG %%%%%%%%%%%%%%%%%%%%%%%%%%%%%%%%%%%%%%%%%%%%%%%%%%%%%%%%%%%

\section{Dependent Types}

In this section, we give examples of proper type dependencies.  First,
we demonstrate that sized types harmonize with predicates, and then,
with large eliminations.

\subsection{Proofs in MiniAgda}

In dependent type theory we represent (co)inductive predicates by
(co)inductive families.  For instance, stream equality, aka stream
bisimilarity, can be defined by the following sized coinductive family.
\begin{samepage}
\begin{verbatim}
sized codata StreamEq (A : Set) : (i : Size) -> Stream A i -> Stream A i -> Set
{ 
  bisim : [i : Size] -> [a : A] -> [as : Stream A i] -> [bs : Stream A i] ->
    StreamEq A i as bs -> 
    StreamEq A ($ i) (cons A i a as) (cons A i a bs)
}
\end{verbatim} % $
\end{samepage}
Two streams are equal if their heads are equal and their tails are
equal.  The latter condition leads to an infinite regression,
therefore, stream equality has to be defined coinductively.  The only
coconstructor $\tbisim$ takes an equality proof of $\vas$ and $\vbs$
to one of $\tcons\;A\;i\;a\;\vas$ and $\tcons\;A\;i\;a\;\vbs$.  The
other arguments of $\tbisim$ are irrelevant, since they can be
reconstructed from the type 
$\StreamEq\;A\;(\$ i)\;(\tcons\;A\;i\;a\;\vas)\;(\tcons\;A\;i\;a\;\vbs)$
(Brady \etal~\cite{bradyMcBrideMcKinna:types03}). 

Note that in the type of the coconstructor $\tbisim$, the size $i$
appears not only as index to the currently defined type $\StreamEq$,
but also to $\Stream$.  This makes sense, since depth $i$ is
sufficient for streams when comparing them up to depth $i$ only.
However, some care is necessary: uses of $i$ need to be restricted in
such a way that $\StreamEq\;A\;i$ is still antitone in $i$.  In our
case, this holds since $\Stream\;A\;i$ itself is antitone in $i$.  If
we replaced $\Stream\;A\;i$ by sized lists $\List\;A\;i$, then
antitonicity of $\StreamEq$ would be lost, leading to unsoundness of
subtyping.  MiniAgda checks types of sized (co)constructors to ensure that
monotonicity properties are retained.

Using $\StreamEq$, we can prove properties about stream functions.
For example, mapping a function $f$ over a stream of $a$s produces a
stream of $(f a)$s.
\begin{code}
\begin{verbatim}
cofun map_repeat : [A : Set] -> [B : Set] -> [i : Size] -> 
  (f : A -> B) -> (a : A) -> 
  StreamEq B i (repeat B (f a) i) (map A B i f (repeat A a i))
{ 
  map_repeat A B ($ i) f a = bisim B i (f a) 
   (repeat B (f a) i) (map A B i f (repeat A a i))
   (map_repeat A B i f a)
}
\end{verbatim}
\end{code}

\subsection{Large Eliminations}

A specific feature of dependent type theory is that one can define a
type by case distinction or recursion on a value, which is called
a large elimination (of the value).  In this case, the
shape of this type (is it a function type or a base type?) is not
statically determined.  Competing approaches to sized types, like $\CIChat$
\cite{bartheGregoirePastawski:lpar06} and $\CACSA$
\cite{blanqui:rta04}, which do not treat sizes explicitly, cannot
define a sized type by a large elimination, because sizes are assigned
automatically and to statically visible (co)inductive types only.

In the following, we define an $n$-ary maximum function on natural
numbers whose type expresses that the size of the output is bounded by the
maximum size of all of the inputs.  A binary maximum function can be
defined as follows, using the built-in $\tmax$ function on sizes.
\begin{samepage}
\begin{verbatim}
fun max2 : [i : Size] -> SNat i -> SNat i -> SNat i
{ max2 i (zero (i > j))    n               = n
; max2 i  m               (zero (i > j))   = m
; max2 i (succ (i > j) m) (succ (i > k) n) = succ (max j k) (max2 (max j k) m n)
}
\end{verbatim}
\end{samepage}
The type of $\tmax2$ expresses that both inputs and the output have
the same upper bound (such a type is beyond $\CIChat$).  By virtue of
subtyping, this type is equivalent to $\SNat\;i \to \SNat\;j \to
\SNat\;(\tmax\;i\;j)$, however, we try to minimize the use of $\tmax$
since constraints like $i \leq \tmax\;j\;k$ slow down
type-checking.\footnote{The constraint $i \leq \tmax\;j\;k$ simplifies
  to the disjunctive constraint $i \leq j \lor i \leq k$ which
  introduces a case distinction in the type checker, possibly
  duplicating computations.}
Now, by induction on $n : \SNat\;\infty$ we define the type
\[\Maxs\;n\;i = \underbrace{\SNat\;i \to \dots \to \SNat\;i}_{\mbox{$n$ times}} \to \SNat\;i\] 
of the $n$-ary maximum function.
\begin{code}
\begin{verbatim}
fun Maxs : SNat # -> Size -> Set
{ Maxs (zero .#  ) i = SNat i
; Maxs (succ .# n) i = SNat i -> Maxs n i
}
\end{verbatim}
\end{code}
The $(n+1)$-ary maximum function $\tmaxs$ is now also defined by
induction on $n$, using the binary maximum.
\begin{code}
\begin{verbatim}
fun maxs : (n : SNat #) -> [i : Size] -> SNat i -> Maxs n i
{ maxs (zero .#)   i m = m
; maxs (succ .# n) i m = \ l -> maxs n i (max2 i m l)
}
\end{verbatim}
\end{code}

\section{Avoiding the Paradoxes}

The theory and implementation of sized types requires some care, since
there are some paradoxes lurking around (as in other areas of
dependent type theory).

\subsection{Non-continuous Types}
\label{sec:adm}

A first paradox noticed by Hughes, Pareto, and Sabry
\cite{hughesParetoSabry:popl96} involves types of recursive functions
which are not continuous in their size parameter.  This phenomenon has
been studied in detail by the author \cite{abel:lmcs07}.  We briefly
explain this issue here for the case of corecursive definitions.

\LONGVERSION{
A divergent program can be constructed which involves a recursive
function of the following type:
\begin{code}
\begin{verbatim}
fun loop : [i : Size] -> SNat i -> (SNat # -> Maybe (SNat i)) -> Unit
\end{verbatim}
\end{code}
The function space in the type $\SNat\;\infty \to \Maybe\;(\SNat\;i)$ of
the last argument destroys continuity in size $i$, leading to a
paradox.  MiniAgda refutes such types of recursive functions
to ensure soundness.
}

To improve readability of the example to follow, let us first
introduce an auxiliary function $\tguard2\;j\;g$ which precomposes 
$g : \Stream\;\Nat\;(\$ j) \to \Stream\;\Nat\;\infty$ 
with itself and with a guard.
\begin{code}
\begin{verbatim}
fun guard2 : [j : Size] -> (Stream Nat ($ j) -> Stream Nat #)
                        -> (Stream Nat j     -> Stream Nat #)
{ guard2 j g xs = g (g (cons Nat j zero xs))
}
\end{verbatim}
\end{code}
We now construct a corecursive function 
$f : [i : \Size] \to (\Stream\;\Nat\;i \to \Stream\;\Nat\;\infty) \to
\Stream\;\Nat\;i$.  It expects an argument $g : \Stream\;\Nat\;i \to
\Stream\;\Nat\;\infty$ whose type promises to add enough guards to a
stream of depth $i$ to make it infinitely guarded.  For any $i <
\omega$, such a function needs to actually add infinitely many guards,
yet for $i = \omega$ it can be any function on streams, even a stream
destructing function like $\ttail$.  This non-continuous behavior
gives rise to a paradox.
\begin{code}
\begin{verbatim}
cofun f : [i : Size] -> (Stream Nat i -> Stream Nat #) -> Stream Nat i
{ f ($ j) g = guard2 j g (f j (guard2 j g)))
}
\end{verbatim}
\end{code}
If $g$ is a stream constructing function or the identity, then
$\tguard2\;j\;g$ is guarding the recursive call to $f$, but if $g$ is
$\ttail$ or another stream destructing function, then $\tguard2\;j\;g$
is actually removing guards.  The definition of $\tf$ is fine as long
as $i$ is not instantiated to $\infty$, otherwise, we can construct a
diverging term.
\begin{code}
\begin{verbatim}
eval let loop : Nat = head Nat # (f # (tail Nat #))
\end{verbatim}
\end{code}
To exclude definitions like $\tf$, MiniAgda checks when matching a
size variable $i$ against a successor pattern $(\$ j)$ that the type
of the result is \emph{upper semi-continuous}
\cite{abel:lmcs07} in $i$.  The check fails since the type
$\Stream\;\Nat\;i \to \Stream\;\Nat\;\infty$ of $g$ is neither antitonic
nor inductive in $i$.

\subsection{Size Patterns and Deep Matching}

The mix of sized types with deep pattern matching leads to a new
paradox which has been communicated to me by Cody Roux.  

Consider the following sized inductive type which has an infinitely
branching $\tL$ and a binary constructor $\tM$.  In the presence of
infinite branching, some sizes are modeled by limit ordinals, and the
closure ordinal is way above $\omega$.  
\begin{code}
\begin{verbatim}
sized data O : Size -> Set
{ Z : [i : Size] -> O ($ i)
; S : [i : Size] -> O i -> O ($ i)
; L : [i : Size] -> (Nat -> O i) -> O ($ i)
; M : [i : Size] -> O i -> O i -> O ($ i)
}
\end{verbatim}
\end{code}
By cases we define a kind of ``predecessor'' function on $\Nat \to
\tO\;(\$\$ i)$ which will be used later to fake a descent.
\begin{code}
\begin{verbatim}
let pre : [i : Size] -> (Nat -> O ($$ i)) -> Nat -> O ($ i)
  = \ i -> \ f -> \ n -> case (f (succ n))
    { (Z .($ i))   -> Z i
    ; (S .($ i) x) -> x
    ; (L .($ i) g) -> g n
    ; (M .($ i) a b) -> a
    } 
\end{verbatim}
\end{code}
The paradox arises when we confuse limit ordinals and successor
ordinals.  In a previous version of MiniAgda, it was possible to write
the following deep match.
\begin{code}
\begin{verbatim}
let three : Nat = succ (succ (succ zero))

fun deep : [i : Size] -> O i -> Nat
{ deep .($$$$ i) (M .($$$ i) (L .($$ i) f)  (S .($$ i) (S .($ i) (S i x))))
  = deep ($$$ i) (M   ($$ i) (L ($ i) (pre i f)) (f three))
; deep i x = zero   
}
\end{verbatim}
\end{code}
We have managed to assign to $f$ the type $\Nat \to \tO\;(\$\$ i)$ on
which we can apply the fake predecessor function.  This makes the
recursive call to $\tdeep$ appear with a smaller size.  It is now easy
to tie the loop:  
\begin{code}
\begin{verbatim}
fun emb : Nat -> O #
{ emb zero = Z #
; emb (succ n) = S # (emb n)
}
eval let loop : Nat = deep # (M # (L # emb) (emb three)) 
\end{verbatim}
\end{code}
In the work of Blanqui \cite{blanqui:rta04}, $\tdeep$ is forbidden by
a linearity condition on size variables (see his Def.~6).  In
MiniAgda, we avoid the paradox by introducing size patterns $(i >
j)$.  Now the left hand side of $\tdeep$ needs to be written as
follows, and the right hand side no longer type-checks.

\pagebreak

\begin{code}
\begin{samepage}
\begin{verbatim}
fun deep : [i : Size] -> O i -> Nat 
{ deep i4 
   (M (i4 > i3) 
        (L (i3 > j2) f) 
        (S (i3 > i2)  
             (S (i2 > i1) 
                  (S (i1 > i) x))))
  = deep _ (M _ (L _ (pre _ f)) (f three))
; deep i x = zero   
}
\end{verbatim}
\end{samepage}
\end{code}  
Now $f : \Nat \to \tO\;j_2$ is no longer a valid argument to $\tpre$
since $j_2$ is not the double-successor of any size expression in scope;
the holes $\_$ cannot be filled with size expressions in a well-typed
manner.

\section{Conclusion}

We have presented MiniAgda, a core dependently typed programming language
with sized types.  The language ensures termination of recursive functions and
productivity of corecursive functions by type checking.  MiniAgda is
implemented in Haskell; the source code and a suite of examples are
available on the author's homepage 
\verb|http://www2.tcs.ifi.lmu.de/~abel/miniagda|.

The main
focus of MiniAgda is the sound integration of sized types with pattern
matching \`a la Agda~\cite{norell:PhD}.  Previous works
\cite{gimenez:typeBased,abel:PhD,hughesParetoSabry:popl96} have treated
sized types in a lambda-calculus with primitives for
fixed points and case distinction.  The exception is the work of Blanqui
\cite{blanqui:rta04,blanqui:csl05} who considers type-based
termination checking of rewrite rules in the Calculus of
Constructions.  Our work is distinguished from his in that we integrate
sizes into the ordinary syntax of dependent types, so, sizes are
first-class.  Also, our treatment includes coinductive types and
productivity. 

In future work, we like to explore how to handle sized types more
silently, such that they are transparent to the user.  There is
already a size constraint solver in MiniAgda which we did not include
in our description.  At the moment, one can replace size arguments on
the right hand side of clauses by ``\_'' and the system searches for
the correct solution.  To complete the picture, we need 
reconstruction of size patterns and a reconstruction of sized types in
function signatures \cite{bartheGregoirePastawski:lpar06}.  
Finally, by adding Agda-style hidden arguments, sizes can disappear
from the surface completely.

\paradot{Acknowledgments} Big thanks to Karl Mehltretter who
implemented the first version of MiniAgda~\cite{mehltretter:diploma}!
Thanks to Ulf Norell with whom I implemented a preliminary version of
sized types for Agda, and thanks to Nils Anders Danielsson for
fruitful discussions on corecursion during the Agda Implementor's
Meetings in Japan in 2008 and 2010.  Thanks to Cody Roux for taking me
to the CORIAS meeting in Val d'Ajol where we had many interesting
discussions about sized types and their implementation.  Finally, I am
grateful for the comments of the anonymous referees that helped me to
improve this paper significantly.

\bibliographystyle{eptcs}%alpha}
\bibliography{auto-eptcs10}

\end{document}

%% file: macros.tex
\newcommand{\der}{\,\vdash}
\def\lv{\mathopen{{[\kern-0.14em[}}}    % opening [[ value delimiter
\def\rv{\mathclose{{]\kern-0.14em]}}}   % closing ]] value delimiter

\newcommand{\dens}[1]{\mathopen{[\kern-0.3ex[}#1\mathclose{]\kern-0.3ex]}}

\def\lo{\mathopen{{\lceil\kern-0.25em\lceil}}}    
\def\ro{\mathclose{{\rfloor\kern-0.25em\rfloor}}}

\newcommand{\abbrev}[1]{#1} % alternative: \emph{#1}

\newcommand{\eg}{\abbrev{e.\,g.}}

\newcommand{\ie}{\abbrev{i.\,e.}}

\newcommand{\etal}{\abbrev{et\,al.}}
\newcommand{\para}[1]{\paragraph*{\it#1}}
\newcommand{\paradot}[1]{\para{#1.}}

\newcommand{\mand}{\ \mbox{and}\ }

\newcommand{\ru}{\dfrac}
\newcommand{\rux}[3]{\ru{#1}{#2}\ #3}

\newcommand{\subst}[3]{#3[#1/#2]}
\newcommand{\SNat}{\mathsf{SNat}}
\newcommand{\tpred}{\mathsf{pred}}
\newcommand{\List}{\mathsf{List}}
\newcommand{\tnil}{\mathsf{nil}}
\newcommand{\tcons}{\mathsf{cons}}
\newcommand{\vas}{\mathit{as}}
\newcommand{\vbs}{\mathit{bs}}
\newcommand{\vxs}{\mathit{xs}}
\newcommand{\Fomega}{\ensuremath{\mathsf{F}^\omega}}
\newcommand{\CIChat}{\ensuremath{\mathsf{CIC}\widehat{~}}}
\newcommand{\CACSA}{\ensuremath{\mathsf{CACSA}}}
\newcommand{\dollar}{\$}
\newcommand{\tminus}{\mathsf{minus}}
\newcommand{\tdiv}{\mathsf{div}}
\newcommand{\Rose}{\mathsf{Rose}}

\newcommand{\tmapRose}{\mathsf{mapRose}}
\newcommand{\tmap}{\mathsf{map}}
\newcommand{\Stream}{\mathsf{Stream}}
\newcommand{\trepeat}{\mathsf{repeat}}

\newcommand{\ttail}{\mathsf{tail}}
\newcommand{\tcase}{\mathsf{case}}
\newcommand{\tof}{\mathsf{of}}
\newcommand{\tmerge}{\mathsf{merge}}
\newcommand{\tfib}{\mathsf{fib}}
\newcommand{\tbad}{\mathsf{bad}}
\newcommand{\tL}{\mathsf{L}}
\newcommand{\tM}{\mathsf{M}}
\newcommand{\tO}{\mathsf{O}}
\newcommand{\tpre}{\mathsf{pre}}
\newcommand{\tdeep}{\mathsf{deep}}
\newcommand{\Maybe}{\mathsf{Maybe}}
\newcommand{\tleq}{\mathsf{leq}}
\newcommand{\tbisim}{\mathsf{bisim}}
\newcommand{\StreamEq}{\mathsf{StreamEq}}
\newcommand{\tmax}{\mathsf{max}}
\newcommand{\Maxs}{\mathsf{Maxs}}
\newcommand{\tmaxs}{\mathsf{maxs}}
\newcommand{\tguard}{\mathsf{guard}}
\newcommand{\tf}{\mathsf{f}}

\newcommand{\Set}{\mathsf{Set}}

\newcommand{\EL}{\mathcal{E}\kern-0.2ex\ell}

%% file: eptcs10.bbl
\begin{thebibliography}{10}
\providecommand{\bibitemstart}[1]{\bibitem{#1}}
\providecommand{\bibitemend}{}
\providecommand{\bibliographystart}{}
\providecommand{\bibliographyend}{}
\providecommand{\url}[1]{\texttt{#1}}
\providecommand{\urlprefix}{Available at }
\providecommand{\bibinfo}[2]{#2}
\bibliographystart

\bibitemstart{abel:PhD}
\bibinfo{author}{Andreas Abel} (\bibinfo{year}{2006}): \emph{\bibinfo{title}{A
  Polymorphic Lambda-Calculus with Sized Higher-Order Types}}.
\newblock \bibinfo{type}{Ph.D. thesis},
  \bibinfo{school}{Ludwig-Maximilians-Universit\"at M\"unchen}.
\bibitemend

\bibitemstart{abel:lmcs07}
\bibinfo{author}{Andreas Abel} (\bibinfo{year}{2008}):
  \emph{\bibinfo{title}{Semi-continuous Sized Types and Termination}}.
\newblock {\sl \bibinfo{journal}{Logical Methods in Computer Science}}
  \bibinfo{volume}{4}(\bibinfo{number}{2}).
\newblock \bibinfo{note}{CSL'06 special issue.}
\bibitemend

\bibitemstart{abel:scp09}
\bibinfo{author}{Andreas Abel} (\bibinfo{year}{2009}):
  \emph{\bibinfo{title}{Type-Based Termination of Generic Programs}}.
\newblock {\sl \bibinfo{journal}{Science of Computer Programming}}
  \bibinfo{volume}{74}(\bibinfo{number}{8}), pp. \bibinfo{pages}{550--567}.
\newblock \bibinfo{note}{MPC'06 special issue.}
\bibitemend

\bibitemstart{abelalti:predicative}
\bibinfo{author}{Andreas Abel} \& \bibinfo{author}{Thorsten Altenkirch}
  (\bibinfo{year}{2002}): \emph{\bibinfo{title}{A Predicative Analysis of
  Structural Recursion}}.
\newblock {\sl \bibinfo{journal}{Journal of Functional Programming}}
  \bibinfo{volume}{12}(\bibinfo{number}{1}), pp. \bibinfo{pages}{1--41}.
\bibitemend

\bibitemstart{DBLP:conf/fossacs/2008}
\bibinfo{editor}{Roberto~M. Amadio}, editor (\bibinfo{year}{2008}):
  \emph{\bibinfo{title}{Foundations of Software Science and Computational
  Structures, 11th International Conference, FOSSACS 2008, Held as Part of the
  Joint European Conferences on Theory and Practice of Software, ETAPS 2008,
  Budapest, Hungary, March 29 - April 6, 2008. Proceedings}}, {\sl
  \bibinfo{series}{Lecture Notes in Computer Science}} \bibinfo{volume}{4962}.
  \bibinfo{publisher}{Springer-Verlag}.
\bibitemend

\bibitemstart{barrasBernardo:fossacs08}
\bibinfo{author}{Bruno Barras} \& \bibinfo{author}{Bruno Bernardo}
  (\bibinfo{year}{2008}): \emph{\bibinfo{title}{The Implicit Calculus of
  Constructions as a Programming Language with Dependent Types}}.
\newblock In \bibinfo{editor}{Amadio}  \cite{DBLP:conf/fossacs/2008}, pp.
  \bibinfo{pages}{365--379}.
\bibitemend

\bibitemstart{gimenez:typeBased}
\bibinfo{author}{Gilles Barthe}, \bibinfo{author}{Maria~J. Frade},
  \bibinfo{author}{Eduardo Gim\'enez}, \bibinfo{author}{Luis Pinto} \&
  \bibinfo{author}{Tarmo Uustalu} (\bibinfo{year}{2004}):
  \emph{\bibinfo{title}{Type-Based Termination of Recursive Definitions}}.
\newblock {\sl \bibinfo{journal}{Mathematical Structures in Computer Science}}
  \bibinfo{volume}{14}(\bibinfo{number}{1}), pp. \bibinfo{pages}{97--141}.
\bibitemend

\bibitemstart{bartheGregoirePastawski:lpar06}
\bibinfo{author}{Gilles Barthe}, \bibinfo{author}{Benjamin Gr{\'e}goire} \&
  \bibinfo{author}{Fernando Pastawski} (\bibinfo{year}{2006}):
  \emph{\bibinfo{title}{{CIC\^{}}: Type-Based Termination of Recursive
  Definitions in the {C}alculus of {I}nductive {C}onstructions}}.
\newblock In: \bibinfo{editor}{Miki Hermann} \& \bibinfo{editor}{Andrei
  Voronkov}, editors: {\sl \bibinfo{booktitle}{Logic for Programming,
  Artificial Intelligence, and Reasoning, 13th International Conference, LPAR
  2006, Phnom Penh, Cambodia, November 13-17, 2006, Proceedings}}, {\sl
  \bibinfo{series}{Lecture Notes in Computer Science}} \bibinfo{volume}{4246},
  \bibinfo{publisher}{Springer-Verlag}, pp. \bibinfo{pages}{257--271}.
\bibitemend

\bibitemstart{bertotKomendantskaya:types08}
\bibinfo{author}{Yves Bertot} \& \bibinfo{author}{Ekaterina Komendantskaya}
  (\bibinfo{year}{2008}): \emph{\bibinfo{title}{Using Structural Recursion for
  Corecursion}}.
\newblock In: \bibinfo{editor}{Stefano Berardi}, \bibinfo{editor}{Ferruccio
  Damiani} \& \bibinfo{editor}{Ugo de'Liguoro}, editors: {\sl
  \bibinfo{booktitle}{Types for Proofs and Programs, International Conference,
  TYPES 2008, Torino, Italy, March 26-29, 2008, Revised Selected Papers}}, {\sl
  \bibinfo{series}{Lecture Notes in Computer Science}} \bibinfo{volume}{5497},
  \bibinfo{publisher}{Springer-Verlag}, pp. \bibinfo{pages}{220--236}.
\bibitemend

\bibitemstart{blanqui:rta04}
\bibinfo{author}{Fr\'ed\'eric Blanqui} (\bibinfo{year}{2004}):
  \emph{\bibinfo{title}{A Type-Based Termination Criterion for
  Dependently-Typed Higher-Order Rewrite Systems}}.
\newblock In: \bibinfo{editor}{Vincent van Oostrom}, editor: {\sl
  \bibinfo{booktitle}{Rewriting Techniques and Applications, 15th International
  Conference, RTA 2004, Aachen, Germany, June 3 -- 5, 2004, Proceedings}}, {\sl
  \bibinfo{series}{Lecture Notes in Computer Science}} \bibinfo{volume}{3091},
  \bibinfo{publisher}{Springer-Verlag}, pp. \bibinfo{pages}{24--39}.
\bibitemend

\bibitemstart{blanqui:csl05}
\bibinfo{author}{Fr{\'e}d{\'e}ric Blanqui} (\bibinfo{year}{2005}):
  \emph{\bibinfo{title}{Decidability of Type-Checking in the {Calculus of
  Algebraic Constructions} with Size Annotations.}}
\newblock In: \bibinfo{editor}{C.-H.~Luke Ong}, editor: {\sl
  \bibinfo{booktitle}{Computer Science Logic, 19th International Workshop, CSL
  2005, 14th Annual Conference of the EACSL, Oxford, UK, August 22-25, 2005,
  Proceedings}}, {\sl \bibinfo{series}{Lecture Notes in Computer Science}}
  \bibinfo{volume}{3634}, \bibinfo{publisher}{Springer-Verlag}, pp.
  \bibinfo{pages}{135--150}.
\bibitemend

\bibitemstart{bradyMcBrideMcKinna:types03}
\bibinfo{author}{Edwin Brady}, \bibinfo{author}{Conor McBride} \&
  \bibinfo{author}{James McKinna} (\bibinfo{year}{2004}):
  \emph{\bibinfo{title}{Inductive Families Need Not Store Their Indices}}.
\newblock In: \bibinfo{editor}{Stefano Berardi}, \bibinfo{editor}{Mario Coppo}
  \& \bibinfo{editor}{Ferruccio Damiani}, editors: {\sl
  \bibinfo{booktitle}{Types for Proofs and Programs, International Workshop,
  TYPES 2003, Torino, Italy, April 30 - May 4, 2003, Revised Selected Papers}},
  {\sl \bibinfo{series}{Lecture Notes in Computer Science}}
  \bibinfo{volume}{3085}, \bibinfo{publisher}{Springer-Verlag}, pp.
  \bibinfo{pages}{115--129}.
\bibitemend

\bibitemstart{coquand:infiniteobjects}
\bibinfo{author}{Thierry Coquand} (\bibinfo{year}{1993}):
  \emph{\bibinfo{title}{Infinite Objects in Type Theory}}.
\newblock In: \bibinfo{editor}{H.~Barendregt} \& \bibinfo{editor}{T.~Nipkow},
  editors: {\sl \bibinfo{booktitle}{Types for Proofs and Programs (TYPES
  '93)}}, {\sl \bibinfo{series}{Lecture Notes in Computer Science}}
  \bibinfo{volume}{806}, \bibinfo{publisher}{Springer-Verlag}, pp.
  \bibinfo{pages}{62--78}.
\bibitemend

\bibitemstart{danielsson:beating}
\bibinfo{author}{Nils~Anders Danielsson} (\bibinfo{year}{2010}):
  \emph{\bibinfo{title}{Beating the Productivity Checker Using Embedded
  Languages}}.
\newblock \bibinfo{note}{Workshop on Partiality And Recursion in Interactive
  Theorem Provers (PAR 2010), Satellite Workshop of ITP'10 at FLoC 2010}.
\bibitemend

\bibitemstart{gimenez:guardeddefinitions}
\bibinfo{author}{Eduardo Gim\'enez} (\bibinfo{year}{1995}):
  \emph{\bibinfo{title}{Codifying Guarded Definitions with Recursive Schemes}}.
\newblock In: \bibinfo{editor}{Peter Dybjer}, \bibinfo{editor}{Bengt
  Nordstr\"{o}m} \& \bibinfo{editor}{Jan Smith}, editors: {\sl
  \bibinfo{booktitle}{Types for Proofs and Programs, International Workshop
  TYPES\'{}94, B\aa{}stad, Sweden, June 6-10, 1994, Selected Papers}}, {\sl
  \bibinfo{series}{LNCS}} \bibinfo{volume}{996}, \bibinfo{publisher}{Springer},
  pp. \bibinfo{pages}{39--59}.
\bibitemend

\bibitemstart{hughesParetoSabry:popl96}
\bibinfo{author}{John Hughes}, \bibinfo{author}{Lars Pareto} \&
  \bibinfo{author}{Amr Sabry} (\bibinfo{year}{1996}):
  \emph{\bibinfo{title}{Proving the Correctness of Reactive Systems Using Sized
  Types}}.
\newblock In: {\sl \bibinfo{booktitle}{23rd ACM SIGPLAN-SIGACT Symposium on
  Principles of Programming Languages, POPL'96}}, pp.
  \bibinfo{pages}{410--423}.
\bibitemend

\bibitemstart{inria:coq82}
\bibinfo{author}{INRIA} (\bibinfo{year}{2008}): \emph{\bibinfo{title}{The Coq
  Proof Assistant Reference Manual}}.
\newblock \bibinfo{organization}{INRIA}, \bibinfo{edition}{version 8.2}
  edition.
\newblock \bibinfo{note}{{http://coq.inria.fr/}}.
\bibitemend

\bibitemstart{jones:sizeChange}
\bibinfo{author}{Chin~Soon Lee}, \bibinfo{author}{Neil~D. Jones} \&
  \bibinfo{author}{Amir~M. Ben-Amram} (\bibinfo{year}{2001}):
  \emph{\bibinfo{title}{The Size-Change Principle for Program Termination}}.
\newblock In: {\sl \bibinfo{booktitle}{ACM Symposium on Principles of
  Programming Languages (POPL'01)}}, \bibinfo{publisher}{ACM Press},
  \bibinfo{address}{London, UK}, pp. \bibinfo{pages}{81--92}.
\bibitemend

\bibitemstart{mcBrideMcKinna:view}
\bibinfo{author}{Conor McBride} \& \bibinfo{author}{James McKinna}
  (\bibinfo{year}{2004}): \emph{\bibinfo{title}{The View from the Left}}.
\newblock {\sl \bibinfo{journal}{Journal of Functional Programming}} .
\bibitemend

\bibitemstart{mehltretter:diploma}
\bibinfo{author}{Karl Mehltretter} (\bibinfo{year}{2007}):
  \emph{\bibinfo{title}{Termination Checking for a Dependently Typed
  Language}}.
\newblock \bibinfo{type}{Master's thesis}, \bibinfo{school}{Department of
  Computer Science, Ludwigs-Maximilians-University Munich}.
\bibitemend

\bibitemstart{miquel:tlca01}
\bibinfo{author}{Alexandre Miquel} (\bibinfo{year}{2001}):
  \emph{\bibinfo{title}{The Implicit Calculus of Constructions}}.
\newblock In: \bibinfo{editor}{Samson Abramsky}, editor: {\sl
  \bibinfo{booktitle}{Typed Lambda Calculi and Applications, 5th International
  Conference, TLCA 2001, Krakow, Poland, May 2-5, 2001, Proceedings}}, {\sl
  \bibinfo{series}{Lecture Notes in Computer Science}} \bibinfo{volume}{2044},
  \bibinfo{publisher}{Springer-Verlag}, pp. \bibinfo{pages}{344--359}.
\bibitemend

\bibitemstart{mishraLingerSheard:fossacs08}
\bibinfo{author}{Nathan Mishra-Linger} \& \bibinfo{author}{Tim Sheard}
  (\bibinfo{year}{2008}): \emph{\bibinfo{title}{Erasure and Polymorphism in
  Pure Type Systems}}.
\newblock In \bibinfo{editor}{Amadio}  \cite{DBLP:conf/fossacs/2008}, pp.
  \bibinfo{pages}{350--364}.
\bibitemend

\bibitemstart{mishraLinger:PhD}
\bibinfo{author}{Richard~Nathan Mishra-Linger} (\bibinfo{year}{2008}):
  \emph{\bibinfo{title}{Irrelevance, Polymorphism, and Erasure in Type
  Theory}}.
\newblock \bibinfo{type}{Ph.D. thesis}, \bibinfo{school}{Portland State
  University}.
\bibitemend

\bibitemstart{norell:PhD}
\bibinfo{author}{Ulf Norell} (\bibinfo{year}{2007}):
  \emph{\bibinfo{title}{Towards a Practical Programming Language Based on
  Dependent Type Theory}}.
\newblock \bibinfo{type}{Ph.D. thesis}, \bibinfo{school}{Department of Computer
  Science and Engineering, Chalmers University of Technology},
  \bibinfo{address}{G\"{o}teborg, Sweden}.
\bibitemend

\bibitemstart{wahlstedt:PhD}
\bibinfo{author}{David Wahlstedt} (\bibinfo{year}{2007}):
  \emph{\bibinfo{title}{Dependent Type Theory with Parameterized First-Order
  Data Types and Well-Founded Recursion}}.
\newblock \bibinfo{type}{Ph.D. thesis}, \bibinfo{school}{Chalmers University of
  Technology}.
\newblock \bibinfo{note}{ISBN 978-91-7291-979-2}.
\bibitemend

\bibitemstart{xi:terminationHOSC}
\bibinfo{author}{Hongwei Xi} (\bibinfo{year}{2002}):
  \emph{\bibinfo{title}{Dependent Types for Program Termination Verification}}.
\newblock {\sl \bibinfo{journal}{Journal of Higher-Order and Symbolic
  Computation}} \bibinfo{volume}{15}(\bibinfo{number}{1}), pp.
  \bibinfo{pages}{91--131}.
\bibitemend

\bibliographyend
\end{thebibliography}
